\documentclass[12pt, a4paper]{article}
\usepackage{graphicx} % Required for inserting images
\usepackage{parskip} %indentationu kaldırıyor
\usepackage{amsmath}
\usepackage{braket} %braket notation
\usepackage{enumitem}
\usepackage{amssymb}
\usepackage{amsthm}
\usepackage{slashed}
\usepackage{wrapfig}
\usepackage{mathtools}
\usepackage[backend=biber,style=nature,doi=true,url=true]{biblatex}
\usepackage{xurl}
\usepackage[hidelinks]{hyperref}

\title{Geometric View of One-Dimensional Quantum Mechanics}
\author{
    Eren Volkan Küçük\\
    \small Department of Physics and Astronomy, Universität Heidelberg, Germany.\\ 
    \small eren.kucuk@stud.uni-heidelberg.de
    }

\date{}

\begin{document}
\setlength{\baselineskip}{1.25\baselineskip}
\setlength{\parskip}{1.25em}

\maketitle

\begin{abstract}
We develop an explicit implementation of De Haro's \textit{Geometric View of Theories} in one-dimensional quantum mechanics. The classical phase space $M = T^*Q$ is taken as the base of a trivial Hilbert bundle $E \cong M \times \mathcal H$, and the familiar position and momentum representations are realised as different trivialisations of this bundle. The Fourier transform appears as a fibrewise unitary transition function, so that the standard position--momentum duality is made precise as a change of coordinates on a single geometric object. For the circle, we also discuss twisted boundary conditions and show how a twist parameter can be incorporated either as a fixed boundary condition or as a base coordinate, in which case it gives rise to a flat $U(\mathcal H)$-connection with nontrivial holonomy. These examples provide a concrete illustration of how the geometric view organises quantum-mechanical representations and dualities in geometric terms.
\end{abstract}

\section{Introduction}

One of the central questions in the philosophy of science concerns the structure of scientific theories: Are they best understood as sets of sentences in a language or as families of models satisfying those sentences \cite{Frigg2022, lutz2014syntactic, halvorson2019logic}? Recent work suggests that both answers are too "flat". They underdescribe the formal and interpretative structure that is actually at work in contemporary physics. The emerging "structured" or "geometric" views of theories insist that additional topological and geometric structure on the space of models is needed, especially once inter-theoretic relations such as dualities, reductions, and approximations are taken seriously \cite{halvorson2019logic, deharo2025geometricviewtheories, deharo2025duality}. In this spirit, De Haro's geometric view of theories (GV) proposes that a physical theory should be seen as an algebraic–geometric object, such as a bundle or manifold equipped with further structure, rather than merely a set of models \cite{deharo2025duality, deharo2025geometricviewtheories}.

Within GV, a particularly concrete realisation is the notion of a \textit{model bundle}, which is a bundle whose fibres collect the models of what, in the more traditional schema, would be called a single theory $T=\langle S,Q,D\rangle$ of states, quantities, and dynamics. Dualities then appear as globally defined transition functions on this bundle, while quasi-dualities appear as transition functions defined only on overlapping regions of the base. In this way, we can say what the underlying geometric object is whose coordinates we are changing, and how the dual descriptions sit as different trivialisations of one and the same bundle.

The aim of this paper is to work out this picture in one of the simplest quantum-mechanical settings: A spinless particle on a circle and a line. This system is standardly used in physics to illustrate basic quantum features such as momentum quantisation, Fourier duality between the position and momentum representations, and the role of boundary conditions. It is therefore a natural test case for the geometric view. If that view cannot be made precise here, it would be hard to trust it in the more demanding cases. At the same time, the system is rich enough to exhibit both exact duality (between position and momentum representations) and mild generalisations (twisted boundary conditions) that resemble quasi-dualities in the sense of De Haro and Butterfield \cite{deharo2025duality}.

A brief remark on the choice of base is in order. Throughout, we take $M=T^*Q$ as a convenient \emph{model space} in the sense of GV \cite{deharo2025duality, deharo2025geometricviewtheories}. It provides a common base admitting both $q$- and $p$-charts over which the trivialisations and transition functions can be formulated. This choice should not be read as a proposal in geometric quantisation; rather, the bundle structure organises equivalent representations of the same quantum system.

The organisation of the paper is as follows: In Section 2, we construct the model bundle for a spinless particle on a line and show how the standard $L^2(\mathbb R)$ and $\ell^2$ descriptions are realised as global trivialisations, related by the Fourier transform viewed as a transition function. In Section 3, we repeat the construction for a particle on a circle, with base $M=T^*S^1$, and in Section 4, we extend this to twisted boundary conditions. There we distinguish two cases: (A) a fixed twist, which affects only the boundary conditions and spectrum; and (B) a twist promoted to a base parameter, where a flat but nontrivial connection encodes the holonomy in the twist direction.  In doing so, the paper illustrates how the duality between position and momentum can be treated, not merely as an abstract isomorphism of Hilbert spaces, but as a genuine change of coordinates on a geometric object. The conclusion summarises what these examples illustrate about the Geometric View and indicates some directions for further work.

We will use standard terminology from fibre bundles and differential geometry. For convenience, a short summary of the needed definitions is given in the appendix.

\section{The Model Bundle for a Spinless Particle on a Line}
Let $M = T^*\mathbb{R} \cong \mathbb{R}^2$, where the cotangent bundle of the configuration space $Q = \mathbb{R}$ with elements $q \in Q$ represents the phase space $T^*\mathbb{R}$ with $p \in T^*Q$, and since it is isomorphic to the 2D Euclidean space, $m = (q,p) \in \mathbb{R}^2$.

Let $\mathcal H$ be an abstract separable Hilbert space. The Hilbert bundle for this example is the trivial bundle
\begin{equation}
  E := M\times \mathcal H,\qquad \pi_E(m,\psi)=m,
\end{equation}
with structure group $U(\mathcal H)$ acting fibrewise. Thus, each fibre is canonically isomorphic to $\mathcal H$:\footnote{See the appendix for more detail.}
\begin{equation}
  \pi_E^{-1}(m)=\{(m,\psi)\mid \psi\in\mathcal H\}\cong\mathcal H.
\end{equation}
Now, let us define two different global trivialisations (since the bundle is trivial) to construct the duality over $M$, as
\begin{equation}
    \begin{split}
        & T_q: \pi_E^{-1}(M) \to \mathbb{R}^2 \times \mathcal{H}(dq) \\
        & T_p: \pi_E^{-1}(M) \to \mathbb{R}^2 \times \mathcal{H}(dp),
    \end{split}
\end{equation}
where $\mathcal{H}(dx) = L^2(\mathbb{R}, dx)$ represents a Hilbert space with the Lebesgue measure $dx$:
\begin{equation}
    L^2(\mathbb{R}, dx) = \left\{ \psi : \mathbb{R} \to \mathbb{C} \; \middle| \; \int_\mathbb{R} |\psi|^2 \; dx < \infty \right\},
\end{equation}
for $x = p \ \text{or} \ x = q$.\footnote{These global trivialisations play a similar role to polarisations in geometric quantisation (GQ) \cite{hurt1983geometric, woodhouse1980geometric}. Apart from this and the bundle structure, there hardly seems to be any similarities between the two frameworks, both in language and purpose, as in GQ, one's goal is to construct a quantum theory from the classical phase space, whereas in GV, the main goal is to construct a geometric model for theories and models.}\textsuperscript{,}\footnote{Here, $\psi \in \mathcal{H}(dx)$ is defined as an equivalence class, $\psi:= [\psi]$ with respect to the equivalence relation $\sim$, under which the functions in $[\psi]$ are the same except at a region of Lebesgue measure zero:
\begin{equation}
     \phi, \psi \in [\psi] = [\phi] \Longleftrightarrow \psi \sim \phi.
\end{equation}
}
We fix the Fourier-transform convention
\begin{equation}
  (\mathcal F \tilde{\psi})(q)
  =\frac{1}{\sqrt{2\pi}}\int_{\mathbb R} e^{iqp}\,\tilde{\psi}(p)\,dp,
  \qquad
  (\mathcal F^{-1} \psi)(p)
  =\frac{1}{\sqrt{2\pi}}\int_{\mathbb R} e^{-iqp}\,\psi(q)\,dq,
\end{equation}
so that $\mathcal F:\mathcal H(dp) \to \mathcal H(dq)$ is unitary. Then, $T_q$ acts as $(m, \psi) \mapsto (m, \psi(q)),$ and $T_p$ as $(m, \psi) \mapsto (m, \tilde{\psi}(p))$. The transition function between the two trivialisations is $h_{qp} := T_q\circ T_p^{-1}$, such that
\begin{equation}
     h_{qp}: \bigl(m, \tilde{\psi}(p) \bigr) \mapsto \bigl(m, \psi(q) \bigr),
\end{equation}
which means, $h_{qp} = (\mathrm{id}_M, \mathcal F) \quad \text{for all } m\in M$. That is, the transition is independent of the base point and acts fibrewise by the Fourier transform. On a suitable dense domain, we have the operators on $\mathcal H(dq)$
\begin{equation}
  (\hat q_q\,\psi)(q)=q\,\psi(q),
  \qquad
  (\hat p_q\,\psi)(q)=-i\,\partial_q \psi(q),
\end{equation}
and on $\mathcal H(dp)$
\begin{equation}
  (\hat p_p\,\tilde{\psi})(p)=p\,\tilde{\psi}(p),
  \qquad
  (\hat q_p\,\tilde{\psi})(p)=i\,\partial_p \tilde{\psi}(p).
\end{equation}
Therefore, we have the intertwining relations
\begin{equation}
  \mathcal F\, (\hat p_p)\, \mathcal F^{-1} = \hat p_q = -i\,\partial_q,
  \qquad
  \mathcal F\, (\hat q_p )\, \mathcal F^{-1} = \mathcal F\, (i\,\partial_p)\, \mathcal F^{-1} = \hat q_q.
\end{equation}
So, the transition function $h_{qp}$ indeed relates the canonical operator representations in the two trivialisations.

This completes the explicit GV data for the line case: A Hilbert bundle over $M$, two global trivialisations corresponding to the $q$- and $p$-descriptions, and a globally defined unitary transition function implementing the familiar position--momentum duality as a change of coordinates on a single geometric object.

\section{The Model Bundle for a Spinless Particle on a Circle}
In this case, the configuration space is a circle, $Q = S^1$, with elements $\theta \in [0, 2\pi), \ \theta \sim \theta + 2\pi$, and radius $R$.\footnote{Although the motivation is different, \textcite{Dolce2023CyclicTime, Dolce2023InternalTimes} also reached similar geometrical results, but considered periodicity in time dimension.} The cotangent space at $\theta$ associated with it is
\begin{equation}
    T_\theta^*S^1 = \{ \alpha \: d\theta \ | \ \alpha \in \mathbb{R} \} \cong \mathbb{R}.
\end{equation}
Therefore, the phase space associated with it is
\begin{equation}
    T^*S^1 \cong S^1 \times \mathbb{R} = \{ (\theta, L) \ | \ \theta \in [0, 2\pi), \ L \in \mathbb{R}  \},
\end{equation}
where $L$ represents angular momentum. That means, the base of the model bundle in this case should be $M = S^1 \times \mathbb{R}$.

The fibre and the structure group stay the same. The PFB and AFB are defined as
\begin{equation}
    \begin{split}
        & P = M \times G = T^*S^1 \times U(\mathcal{H}), \\
        & E = P \times_G F = (T^*S^1 \times U(\mathcal{H})) \times_{U(\mathcal{H})} \mathcal{H} \equiv (S^1\times \mathbb{R}) \times \mathcal{H},
    \end{split}
\end{equation}
where $(m, \psi) = \bigl((\theta, L), \psi \bigr) \in E$. The projection stays as defined above.

To define the trivialisations, we need to look at the system a little closer: In the position representation, the time-independent Schrödinger equation is\footnote{Although we only need the boundary conditions and not the dynamics, we started with the Schrödinger equation for the sake of completeness.}
\begin{equation}
    \frac{-\hbar^2}{2m_0R^2}\partial^2_\theta\phi(\theta) + V(\theta)\phi(\theta) = E \phi(\theta),
\end{equation}
where $m_0$ is the mass of the particle, and $R$ is the radius of the circle. This is an eigenvalue problem,
\begin{equation}
    \hat{H}\phi_n(\theta) = E_n\phi_n(\theta),
\end{equation}
with periodic boundary conditions (twisted boundary conditions are considered in the next section)
\begin{equation}
\label{eq: periodic-bc}
    \phi_n(\theta) = \phi_n(\theta + 2\pi), \\ n \in \mathbb{Z},
\end{equation}
which yields a complete orthonormal basis $\{\phi_n\}$, for the general solution
\begin{equation}
    \psi(\theta) = \sum_{n \in \mathbb{Z}} c_n \phi_n(\theta).
\end{equation}
The coefficients $c_n$ and $\psi(\theta)$ are related by the Fourier transform
\begin{equation}
    c_n = \int_0^{2\pi} \psi(\theta) e^{-in\theta} \frac{d\theta}{2\pi} \Longleftrightarrow \psi(\theta) = \sum_{n \in \mathbb{Z}} c_n e^{i n\theta}.
\end{equation}
Therefore, the position representation of the Hilbert space is
\begin{equation}
     L^2(S^1, d\theta/2\pi) = \left\{ \psi : S^1 \to \mathbb{C} \; \middle| \; \int_0^{2\pi} |\psi(\theta)|^2 \; d\theta/2\pi < \infty \right\},
\end{equation}
with the inner product
\begin{equation}
  \langle \psi,\phi\rangle
  \;=\;
  \int_0^{2\pi}\overline{\psi(\theta)}\,\phi(\theta)\,\frac{d\theta}{2\pi}.
\end{equation}
The momentum representation is
\begin{equation}
    \ell^2(\mathbb{Z}) = \left\{ c =  (c_n) \; \middle| \; c_n : \mathbb{Z} \to  \mathbb{C}, \ \sum_{n \in \mathbb{Z}} |c_n|^2 < \infty \right\},
\end{equation}
with the inner product
\begin{equation}
    \braket{c, d} = \sum_{n \in \mathbb{Z}} \bar{c}_n d_n < \infty.
\end{equation}
The trivialisations, which are again global, thus become
\begin{equation}
  \begin{split}
    &T_\theta : \pi_E^{-1}(M) \longrightarrow
      (S^1\times\mathbb R)\times L^2(S^1,d\theta/2\pi),\\[2mm]
    &T_L : \pi_E^{-1}(M) \longrightarrow
      (S^1\times\mathbb R)\times \ell^2(\mathbb Z).
  \end{split}
\end{equation}
For each base point $m\in M$, the transition function, $h_{\theta L}:= T_\theta \: \circ \: T_L^{-1}$, is a unitary operator on the fibre $\pi^{-1}_E(m) \cong\mathcal H$, which relates the same abstract state to its two representations, that is, momentum coefficients $c\in\ell^2(\mathbb Z)$ and position wave-function $\psi(\theta)\in L^2(S^1,d\theta/2\pi)$.\footnote{A (smooth) section of the Hilbert bundle $\pi_E:E\to M$ is a map
\begin{equation}
  s : M \longrightarrow E
  \qquad\text{with}\qquad
  \pi_E\circ s = \mathrm{id}_M.
\end{equation}
In the present trivial situation $E \cong M\times\mathcal H$, a section can be written as
\begin{equation}
    s(m) = (m,\psi(m)),
\end{equation}
for some map $\psi:M\to\mathcal H$. Thus, a section is a choice, for each phase space point $m\in M$, of a state vector in the corresponding fibre. In particular, a fixed state $\psi\in\mathcal H$ corresponds to the constant section $s_\psi(m) = (m,\psi)$. When we speak of “wave-functions” in what follows, we mean such local representatives $\psi(\cdot)$ of sections in a chosen trivialisation. For instance, in the position trivialisation \(T_\theta\), this reads
\begin{equation}
    s(m) = (m, \psi(\theta; m)),
\end{equation}
where \(\psi(\theta; m) \in L^2(S^1)\) may depend on \(m = (\theta, L)\).} Concretely:
\begin{equation}
  h_{\theta L} : (m,c) \longmapsto \bigl(m,\psi(\theta)\bigr),
  \qquad
  \psi(\theta) = \sum_{n\in\mathbb Z} c_n e^{in\theta}.
\end{equation}
Thus $h_{\theta L}$ acts fibrewise, that is, as the identity on the base point $m$ and as the Fourier transform on the fibre.

\section{Twisted Boundary Conditions on $M = T^*S^1$}
As an example of a more interesting system, we are going to analyse systems with twisted boundary conditions. These kinds of systems are introduced in the seminal papers \cite{aharonov1959significance, berry1984phase}.\footnote{To learn more about the mathematics of the subject, see chapters 16 and 17 in \cite{frankel2012geometry}, and 10 in \cite{nahakara2003geometry}.}

As in the previous section, we are going to consider a particle on a circle, with coordinate $\theta\in[0,2\pi)$ and $\theta\sim\theta+2\pi$. Instead of strict periodicity, we allow the twisted boundary condition
\begin{equation}
  \psi(\theta+2\pi) \;=\; e^{i\alpha}\,\psi(\theta),
  \qquad \alpha\in\mathbb{R}.
  \label{eq:twisted-BC}
\end{equation}
We distinguish two situations:
\begin{enumerate}
  \item[(A)] $\alpha$ is a constant,
  \item[(B)] $\alpha$ is treated as a classical parameter.
\end{enumerate}

\subsection{Case A: Fixed Twist \texorpdfstring{$\alpha$}{alpha}}

In Case~A, the twist $\alpha$ is fixed. The base is purely classical,
\begin{equation}
      M \;=\; T^*S^1 \;\cong\; S^1 \times \mathbb{R},
\end{equation}
and $\alpha$ is not a coordinate on $M$.

For fixed $\alpha$, the position representation is
\begin{equation}
\begin{split}
      \mathcal{H}^{(\alpha)}_\theta := \left\{\psi:[0,2\pi)\to\mathbb{C} \: \middle| \: \psi(\theta+2\pi)=e^{i\alpha}\psi(\theta), \int_0^{2\pi}|\psi(\theta)|^2\,\frac{d\theta}{2\pi}<\infty \right\},
  \label{eq:H-theta-alpha}
\end{split}
\end{equation}
with inner product
\begin{equation}
  \langle \psi,\phi\rangle
  \;=\;
  \int_0^{2\pi}\overline{\psi(\theta)}\,\phi(\theta)\,\frac{d\theta}{2\pi}.
\end{equation}
Writing $\varphi:=\alpha/(2\pi)$, the shifted modes
\begin{equation}
  e^{i(n+\varphi)\theta},
  \qquad n\in\mathbb{Z},
\end{equation}
satisfy the boundary condition \eqref{eq:twisted-BC} and form an orthonormal basis of $\mathcal{H}^{(\alpha)}_\theta$. On this basis, the angular momentum operator
\begin{equation}
      \hat L \;=\; -\,i\hbar\,\partial_\theta
\end{equation}
has the eigenvalues
\begin{equation}
  \hat L\,e^{i(n+\varphi)\theta}
  \;=\;
  \hbar (n+\varphi)\,e^{i(n+\varphi)\theta},
  \qquad n\in\mathbb{Z},
\end{equation}
so the spectrum is $L=(n+\varphi)\hbar$. For the free particle of mass $m_0$ on a circle
of radius $R$, the corresponding energy levels are
\begin{equation}
  E_n \;=\; \frac{\hbar^2}{2m_0R^2}\,(n+\varphi)^2.
\end{equation}
The momentum representation is the sequence space of the previous case,
\begin{equation}
  \mathcal{H}_L
  \;:=\;
  \bigl\{ c=(c_n)_{n\in\mathbb{Z}} \,\big|\, \sum_{n\in\mathbb{Z}}|c_n|^2<\infty \bigr\}
  \;\cong\; \ell^2(\mathbb{Z}),
\end{equation}
with inner product $\langle c,d\rangle = \sum_{n\in\mathbb{Z}}\overline{c_n}d_n$.
The AFB $E$ is trivial again, as
\begin{equation}
      E \;\cong\; M \times \mathcal H.
\end{equation}
Now, we choose two global trivialisations:
\begin{align*}
  T_L &: E \to M \times \mathcal{H}_L,\\
  T_\theta^{(\alpha)} &: E \to M \times \mathcal{H}^{(\alpha)}_\theta.
\end{align*}
The transition function, $h_{\theta L}:= T^{(\alpha)}_\theta \: \circ \: T_L^{-1}$, between these frames is the constant map
\begin{equation}
  h_{\theta L}: (m,c) \longmapsto \bigl(m,\psi(\theta)\bigr),
  \quad
  (\mathcal{F}_\alpha c)(\theta) = \psi(\theta),
  \quad \forall m\in M,
\end{equation}
where $\mathcal{F}_\alpha:\mathcal{H}_L\to\mathcal{H}^{(\alpha)}_\theta$ is the
shifted Fourier transform
\begin{equation}
  (\mathcal{F}_\alpha c)(\theta)
  \;=\;
  \sum_{n\in\mathbb{Z}} c_n\,e^{i(n+\varphi)\theta}.
  \label{eq:F-alpha}
\end{equation}
Thus $h_{\theta L}$ acts trivially on the base point $m$ and unitarily on the fibre.

As it is seen, apart from an extra constant factor $\alpha \in \mathbb{R}$, there is essentially not much of a difference between the boundary conditions
\eqref{eq: periodic-bc} and \eqref{eq:twisted-BC}. The following case is more interesting in that respect.

\subsection{Case B: The Twist $\alpha$ as a Base Parameter}
In Case~B, we promote $\alpha$ to a classical parameter and include it in the base,
\begin{equation}
  M' \;=\; T^*S^1 \times S^1_\alpha
  \;\cong\;
  (S^1 \times \mathbb{R}) \times S^1_\alpha \cong T^2 \times \mathbb{R},
\end{equation}
where $S^1_\alpha$ parametrises $\alpha$ modulo $2\pi$. A point of $M'$ is $m'= (m, \alpha) = (\theta,L,\alpha)$.

For each $\alpha$, the position space is $\mathcal{H}^{(\alpha)}_\theta$ as in
eq.\eqref{eq:H-theta-alpha}. As $\alpha$ varies, we thus obtain a family of
Hilbert spaces
\begin{equation}
      \{\mathcal{H}^{(\alpha)}_\theta\}_{\alpha\in S^1_\alpha},
\end{equation}
all of which are unitarily isomorphic to one another, but with different twisted boundary conditions \eqref{eq:twisted-BC}. On the other hand, the momentum space $\mathcal{H}_L\cong\ell^2(\mathbb{Z})$
is independent of $\alpha$. 

We again take $E'\to M'$ to be trivial and use two trivialisations
\begin{align*}
  T_L &: E' \to M'\times\mathcal{H}_L,
  \\
  T_\theta &: E' \to M'\times\mathcal{H}^{(\alpha)}_\theta,
\end{align*}
where in the second line we emphasise the $\alpha$-dependence of the position space. The transition function is now $\alpha$-dependent:
\begin{equation}
  h_{\theta L}(m,\alpha)
  \; = \;
  \mathcal{F}_\alpha:\ \mathcal{H}_L \to \mathcal{H}^{(\alpha)}_\theta,
  \qquad (m,\alpha)\in M',
\end{equation}
with $\mathcal{F}_\alpha$ still given by eq.\eqref{eq:F-alpha} but now viewed as a smooth family of unitary operators parametrised by $\alpha$. Note that $\mathcal F_\alpha$ is not strictly periodic in $\alpha$:
\begin{equation}
  \mathcal F_{\alpha+2\pi} = M_{e^{i\theta}}\,\mathcal F_\alpha,
\end{equation}
where $M_{e^{i\theta}}$ denotes multiplication by $e^{i\theta}$ on $\mathcal H_\theta^{(\alpha)}\subset L^2(S^1,d\theta/2\pi)$. Thus, when $\alpha$ is treated as a coordinate on $S^1_\alpha$, the family closes only up to a nontrivial gauge transformation.

Let us choose the momentum frame so that the connection is flat and trivial:
\begin{equation}
  A_L \;=\; 0,
  \qquad F_L \;=\; 0.
\end{equation}
Transforming to the position frame (see eq.\eqref{eq:A-transform} in the appendix) gives
\begin{equation}
  A_\theta
  \;=\;
  -\,(d\mathcal{F}_\alpha)\,\mathcal{F}_\alpha^{-1}.
\end{equation}
Here $d$ includes differentiation with respect to $\alpha$, since
$\mathcal{F}_\alpha$ depends on $\alpha$. A short calculation using
\begin{equation}
    \frac{\partial}{\partial\alpha}
  e^{i\bigl(n+\alpha/2\pi\bigr)\theta}
  \;=\;
  i\,\frac{\theta}{2\pi}\,
  e^{i\bigl(n+\alpha/2\pi\bigr)\theta}
\end{equation}
shows that $A_\theta$ acts, in the position frame, as multiplication by
$i\theta/(2\pi)$ along the $\alpha$-direction:
\begin{equation}
  A_\theta \psi(\theta;\alpha)
  \;=\;
  i\,\frac{\theta}{2\pi}\,\psi(\theta;\alpha)\,d\alpha.
\end{equation}
Thus, the covariant derivative along $\alpha$ is
\begin{equation}
  D_\alpha \psi
  \;=\;
  \frac{\partial\psi}{\partial\alpha}
  \;+\;
  i\,\frac{\theta}{2\pi}\,\psi.
\end{equation}
The curvature remains zero,
\begin{equation}
  F_\theta
  \;=\;
  dA_\theta + A_\theta\wedge A_\theta
  \;=\;
  0,
\end{equation}
so the connection is flat but has nontrivial holonomy along the
$\alpha$-circle: Transporting a state once around
\begin{equation}
    \alpha:0\to 2\pi
\end{equation}
at fixed $(\theta,L)$ yields
\begin{equation}
  \psi(\theta;\alpha)
  \longmapsto
  \exp\!\left(\int_0^{2\pi} i\,\frac{\theta}{2\pi}\,d\alpha\right)\psi(\theta;\alpha)
  \;=\;
  e^{i\theta}\,\psi(\theta;\alpha).
\end{equation}
This is holonomy in the parameter-space direction $\alpha$. It arises only in Case~B, where $\alpha$ is part of the base and can vary. Topologically, $E'$ remains trivial as a Hilbert bundle. The nontriviality resides in the connection, not in the bundle’s isomorphism class.

The momentum representation stays $\ell^2(\mathbb{Z})$ in all cases. The labels of position-basis plane waves are shifted by $\varphi=\alpha/2\pi$, yielding the spectrum $L=(n+\varphi)\hbar$ and $E_n=\frac{\hbar^2}{2m_0R^2}(n+\varphi)^2$ for the free particle.

\section{Conclusion}

In this paper, we have provided a fully explicit realisation of the geometric view \cite{deharo2025duality, deharo2025geometricviewtheories} in elementary quantum mechanics, using the spinless particle on a line and on a circle as test cases. The central idea was to treat the familiar position and momentum representations not merely as abstractly isomorphic Hilbert spaces, but as different global frames (trivialisations) of a single Hilbert bundle over a common model space. In this formulation, the Fourier transform appears as a fibrewise unitary transition function, so that the standard position--momentum duality is made precise as a change of coordinates on one underlying geometric object.

For the particle on a circle, the same bundle-theoretic picture captures momentum quantisation through the discrete Fourier transform between $L^2(S^1)$ and $\ell^2(\mathbb Z)$, again interpreted as a transition between global frames. The twisted case then illustrates a further aspect that is central to GV: The way in which parameters may be treated either as a fixed background structure or as part of the base of the model bundle. When the twist $\alpha$ is held fixed, it is encoded entirely in the choice of twisted position space and in the shifted Fourier basis, leading to the familiar spectral shift $L=(n+\alpha/2\pi)\hbar$ for the angular momentum. When $\alpha$ is instead promoted to a base coordinate, the resulting $\alpha$-dependent family of frames induces a flat $U(\mathcal H)$-connection whose curvature vanishes, but whose holonomy around the $\alpha$-circle is nontrivial. In this way, the examples make transparent how GV separates local equivalence (flatness, vanishing curvature) from global structure (holonomy), and how global information may persist even when the local dynamics is unchanged.

Methodologically, these models support the guiding claim of GV that the structure of a physical theory is not exhausted by a set of models or sentences, but includes additional geometric data. The bundle formulation makes it clear what is being identified across representations (abstract fibre states) and what is being changed (the frame/trivialisation), and it provides a natural language for describing cases where transition maps fail to be globally single-valued without a compensating gauge transformation.

Several extensions suggest themselves. First, one can move beyond the trivial bundle setting and consider systems where only local trivialisations exist. Second, one may treat other parameters (such as couplings, background fields, or boundary data) as base directions and study the induced connections and holonomies as invariants of families of representations. Finally, the present constructions can serve as templates for more complex examples (spin, external gauge fields, many-body systems), where representation change and duality are richer and where GV's organising role may be even more valuable.

Overall, the line and circle examples show that GV can be made completely explicit in ordinary quantum mechanics and that doing so yields a clear geometric account of representation change, duality, and parameter dependence, which are precisely the kinds of structure that the Geometric View was designed to capture.

\appendix \section{Definitions and Conventions for Fibre Bundles and Differential Geometry}
Throughout the text, we used the language of fibre bundles and differential geometry, so let us introduce some definitions.\footnote{For more on the fibre bundles, differential geometry, and their uses in physics, see \cite{hassani2013mathematical, nahakara2003geometry, frankel2012geometry}.} 

\subsection*{Fibre Bundles}
Let $M$ be the base, $G$ a (Lie) group, and $\mathcal{H}$ a separable Hilbert space with a left unitary representation
\begin{equation}
    \rho: G \to U(\mathcal{H}), \qquad (g,\psi)\mapsto \rho(g)\psi \equiv g\psi.
\end{equation}
Let $P \to M$ be any principal $G$-bundle with right action $P\times G\to P$, $(p,g)\mapsto p\cdot g$.
Define an equivalence relation on $P\times\mathcal{H}$ by
\begin{equation}
\label{pg, f - p, gf}
  (p\cdot g,\ \psi)\ \sim\ (p,\ g\psi)\qquad
  \text{for all } p\in P,\ g\in G,\ \psi\in\mathcal{H}.
\end{equation}
The associated Hilbert bundle is then
\begin{equation}
  E \;:=\; (P\times\mathcal{H})\big/\!G,
  \qquad [p,\psi]\in E,
  \qquad \pi_E([p,\psi]) := \pi_P(p).
\end{equation}
In the main text, the fibre $F$ is the abstract Hilbert space, $\mathcal{H}$, defined over the field of complex numbers with an inner product\footnote{See \cite{cohen2020quantum} to learn more about the quantum mechanical parts in the paper in more detail.}
\begin{equation}
    \braket{\cdot , \cdot} : \mathcal{H} \times \mathcal{H} \to \mathbb{C},
\end{equation}
where $\psi \in \mathcal{H}$. The structure group $G$ of the bundle is the unitary group 
$$U(\mathcal{H}) = \left\{ h: \mathcal{H} \to \mathcal{H} \; | \; \braket{h \psi, h \phi} = \braket{\psi, \phi}, \;\; \forall \phi, \psi \in \mathcal{H}  \right\},$$
where the elements $h$ also satisfy linearity and are bijective over $\mathcal{H}$. As usual, $h \in U(\mathcal{H})$ act $\mathcal{H}$ on the left. The principal fibre bundle (PFB) associated with it is trivial
\begin{equation}
    P = M \times G.
\end{equation}
Therefore, the associated fibre bundle (AFB) 
$$E = (P \times F)/G = (\mathbb{R}^2 \times U(\mathcal{H})) \times_{U(\mathcal{H})} \mathcal{H} $$ 
is automatically trivial, as
\begin{equation}
\label{equivalence of E}
    E = (\mathbb{R}^2 \times U(\mathcal{H})) \times_{U(\mathcal{H})} \mathcal{H} \equiv \mathbb{R}^2 \times \mathcal{H},
\end{equation}
where $m  \in \mathbb{R}^2$, $h \in U(\mathcal{H})$, and $\psi \in \mathcal{H}$. Here $G=U(\mathcal H)$ acts on $P\times F$ by
\begin{equation}
  g\cdot\bigl((m,h),\psi\bigr)
  := \bigl((m,hg^{-1}),\,g\psi\bigr),
  \qquad g,h\in U(\mathcal H),\ \psi\in\mathcal H,
\end{equation}
and $[(m,h),\psi]$ denotes the corresponding equivalence class in $E$,
\begin{equation}
    [(m,h), \psi] \in (\mathbb{R}^2 \times U(\mathcal{H})) \times_{U(\mathcal{H})} \times \mathcal{H},
\end{equation}
where $(m,h) \in \mathbb{R}^2 \times U(\mathcal{H})$. The logic behind the equivalence in eq.\eqref{equivalence of E} is that every class admits a canonical representative with principal part \((m,\mathrm{id})\), for
\begin{equation}
    [\,(m,h),\ \psi\,]\;=\;\bigl[(m,\mathrm{id}),\ h\psi\bigr]
\end{equation}
by the definition in eq.\eqref{pg, f - p, gf}.\footnote{That is, the elements $(m,h, \psi)$ and $(m, id, h\psi)$ are both in the same set, which can be denoted as $[(m,h), \psi]$ or $[(m,\text{id}), h\psi]$.} Therefore, we can set\footnote{That means, for every element $m \in M$, the element $[(m, h), \psi] \in E$ can be represented by the element with the identity element of the group $G$. Therefore, we can set the equivalence relation
\begin{equation}
    \equiv: E \to \mathbb{R}^2 \times \mathcal{H}
\end{equation}
such that
\begin{equation}
    \equiv: [(m, h), \psi] = [(m, \text{id}), h\psi] \mapsto (m, \psi)
\end{equation}
or just
\begin{equation}
    [(m, \text{id}), h\psi] \equiv (m, \psi),
\end{equation}
from which the triviality of AFB $E$ follows. 
}
\begin{equation}
    [(m, h), \psi] = [(m, \text{id}), h\psi] \equiv (m, \psi). 
\end{equation}
The projection $\pi_E: E \to M$ maps elements in $E$ as $(m, \psi) \mapsto m$. The fibres are defined as $\pi_E^{-1}(m) = \{ (m, \psi) \in E \ | \psi \in \mathcal{H} \ \} \cong \mathcal{H}.$ So, we have a copy of the Hilbert space over every point of the phase space $M = \mathbb{R}^2$.

\subsection*{Differential Geometry}
As is done in the main text, let $E$ be a Hilbert bundle with typical fibre $\mathcal{H}$ and base $M$. On overlaps $U\cap V\neq\varnothing$, where $U,V \subset M$, the trivialisations are related by a transition function
\begin{equation}
  g_{UV}: U\cap V \longrightarrow U(\mathcal{H}),
  \qquad
  T_V \circ T_U^{-1}(m,\psi)
  \;=\;
  \bigl(m,\,g_{UV}(m)\psi\bigr).
\end{equation}
Now, a connection on a vector (or Hilbert) bundle $E$ is, by definition, an operator that differentiates sections of $E$, i.e.\ smooth choices of a vector in the fibre over each base point. Formally, a section is a map
\begin{equation}
    s: M \longrightarrow E
  \qquad\text{such that}\qquad
  \pi_E\circ s = \mathrm{id}_M,
\end{equation}
so in a trivialisation $E \simeq M\times\mathcal H$ we can write
\begin{equation}
      s(m) = (m,\psi(m)),
  \qquad \psi: M \to \mathcal H.
\end{equation}
A connection $D$ is then a map
\begin{equation}
      D: \Gamma(E) \longrightarrow \Omega^1(M;E),
\end{equation}
which locally takes the form
\begin{equation}
      D\psi = d\psi + A\,\psi,
\end{equation}
where $\psi:U\to\mathcal H$ is the local representative of a section,
$d\psi$ is its ordinary differential, and $A$ is an operator-valued
one-form. Thus, connections and their curvature naturally act on sections. They describe how a state assigned to each point of $M$ changes as we move in the base. In our trivial Hilbert bundle $E \simeq M\times\mathcal H$, a fixed state $\psi\in\mathcal H$ corresponds to the constant section $s_\psi(m)=(m,\psi)$, so we will not distinguish notationally between $\psi\in\mathcal H$ and the associated section.

A covariant derivative,
\begin{equation}
  D\psi \;=\; d\psi + A_U\,\psi,
  \qquad \psi:U\to\mathcal{H},
\end{equation}
with an operator-valued 1-form,
\begin{equation}
    A_U \;\in\; \Omega^1(U \subset M;\,\mathfrak u(\mathcal{H})),
\end{equation}
defines a curvature as
\begin{equation}
  F_U \;:=\; dA_U + A_U\wedge A_U \;\in\; \Omega^2(U;\,\mathfrak u(\mathcal{H})),
\end{equation}
where $\mathfrak u(\mathcal{H})$ is the Lie algebra associated with the group $U(\mathcal{H})$. On overlaps $U\cap V \neq \emptyset$, $(A_U,F_U)$ and $(A_V,F_V)$ are related by the transition functions according to
\begin{align}
  A_V
  &=\;
  g_{UV}\,A_U\,g_{UV}^{-1} \;-\; (dg_{UV})\,g_{UV}^{-1},
  \label{eq:A-transform}
  \\
  F_V
  &=\;
  g_{UV}\,F_U\,g_{UV}^{-1}.
\end{align}
These transformation rules express the fact that $D$ and $F$ are globally
well-defined, even though the local representatives $A_U$ depend on the choice of trivialisation.

\small \textbf{Statements and Declarations}

\footnotesize{ \textbf{Competing Interests}  
The author declares that there are no financial or non-financial competing interests related to this work.

\textbf{Funding}  
No funding was received for conducting this study.

\textbf{Data Availability} No datasets were generated or analysed during the current study.
}

\printbibliography

\end{document}